# Weighted-Interaction Nestedness Estimator (WINE): A new estimator to calculate over frequency matrices

**Software availability**

Name of software: WINE

Developers: Javier Galeano, Juan M. Pastor, and Jose M. Iriondo.

Platform: Matlab

Contact address: Dept. Ciencia y Tecnología Aplicadas a la I.T. Agrícola, E.U.I.T. Agrícola. Universidad Politécnica de Madrid, Ciudad Universitaria s/n, E-28040 Madrid, Spain.

Telephone: +34 913365445

E-mail address: javier.galeano@upm.es

Availability: http://hypatia.agricolas.upm.es/WINE/

A common working procedure in Ecology is to identify patterns and elaborate hypotheses about the processes that may be responsible for their occurrence (Levin, 1992). One approach to identifying and characterizing patterns in community and network ecology is through the calculation of nestedness (Bascompte et al., 2003). Nestedness is a measure of a particular type of pattern in an ecological system, referring to the order that emanates from the way elements of a particular set are linked to elements of a second set. These links may relate, for instance, to the interactions that are established between two sets of species in an ecosystem (e.g., plant-pollinator interactions, Bascompte et al., 2003), or to the occurrence of a set of species in a given set of habitat fragments of different sizes (Atmar and Patterson, 1993). Thus, in the latter case, species assemblages are nested if the species present in species-poor sites are proper subsets of the assemblages found in species-rich sites (Patterson and Atmar, 1986), and perfect nestedness occurs when all species-poor sites are proper subsets of the assemblages found in richer-species sites (Almeida-Neto et al., 2007). It should be noted, though, that absence of nestedness does not always mean absence of pattern. Several other types of patterns, such as gradients and compartments, may also be found in ecological systems (Leibold and Mikkelson, 2002; Lewinsohn et al., 2006; Almeida-Neto et al., 2007). Nestedness can be assessed through an ordered binary presence-absence matrix in which nestedness leads to a maximally packed pattern of ones and zeros (Ulrich and Gotelli, 2007). Thus, unexpected presences or absences from a maximally packed matrix can be used to quantify nestedness (Atmar and Patterson, 1993). Although nestedness has its origins in ecology, this concept and its extensions are interesting properties of datasets that can also be applied to many other domains (Mannila and Terzi, 2007; May, 2008)

During the last years several different approaches have been proposed for estimating nestedness and, accordingly, several programs have been created and made available to calculate nestedness indices (Ulrich et al., 2009). The most popular approach for estimating nestedness, based on the matrix's temperature, $T$, was the Nestedness Temperature Calculator, (NTC) (Atmar and Patterson, 1993). This Visual Basic software calculates temperature as a measure of how much the presence-absence matrix departs from perfect nestedness. Guimarães and Guimarães (2006) published a C++ software, called ANINHADO. This program automatically obtains the same temperature, $T$, over 10,000 matrices and generates empirical distributions for four pre-defined null models. Rodríguez-Gironés and Santamaría (2006) presented the program BINMATNEST, introducing a new algorithm to overcome some weaknesses found in Atmar and Patterson's temperature (dependency of matrix size and fill, and packing algorithm). This program uses genetic algorithms for packing and provides three alternative null models for calculating the statistical significance of matrix temperature. In a broad review of nestedness indices, Ulrich and Gotelli (2007) showed that NTC and its modified versions did not perform well as an index for detecting nestedness. Instead, Brualdi and Sanderson's discrepancy index (Brualdi and Sanderson, 1999) and Cutler's index of unexpected presences (Cutler, 1991) provided the best results. Recently, Corso et al. (2008) and Almeida-Neto et al. (2008) have defined new ways to calculate nestedness. Corso and coworkers estimate nestedness through a rescaled Manhattan distance, whereas NODF, the metric of Almeida-Neto et al., is based on the comparison of pairs of columns or rows using to properties associated to nestedness: decreasing fill and overlap.

All these precedent estimators and corresponding software programs use presence-absence adjacency matrices (binary matrices) as the basis for calculating nestedness. This simple approach facilitates the description and characterization of the topology of the network. However, networks are specified not only by their topology but also by the heterogeinity in the weight (or the intensity) of the connections (Barrat *et al.*, 2004). Characterizing links with presence-absence data does not take into account the possible differences in intensity among links. For example, when studying plant-pollinator interactions, a pollinator may visit the flowers of a particular plant species 70% of the times but only occasionally visit the flowers of other plant species (e.g., less than 5% of the times in each species). In such a case, the classical presence-absence would treat all these links similarly (i.e., as presences). In this paper we introduce a new nestedness estimator that takes into account the weight of each interaction. Defining an "event" as a registered occurrence that is used to define a link, the weight represents the total number of events recorded for a particular link. For example, in a plant-pollinator network, a link is established when a pollinator species visits a plant species. The weight of this link is the number of registered visits of this kind. Instead of using presence-absence matrices, we calculate the new estimator from quantitative data matrices that include the number of events of each interaction, such as the number of visits in the plant-pollinator case. Thus, this is the first estimator that allows for the characterization of weighted nestedness.

To calculate the weighted-interaction nestedness estimator, we start with the matrix containing the number of events of each interaction, $M_{ij}$. The matrix is then packed in the following way: the row with the greatest marginal totals (i.e., number of links) is rearranged as the last row. Subsequently, the next row with the greatest marginal totals is placed in the previous row, and so on, until all the elements assigned to rows are arranged in descending order from the last to the first row according to the marginal totals. Similarly, the elements of the set assigned to columns are arranged so that the column with the greatest marginal totals is rearranged in the rightmost column and the rest in descending order toward the left. Thus, nestedness will be related to the proximity of existing links to one another in the packed matrix, so that the most nested matrix will be the one that after packing shows a minimum mixing of filled cells (links) with empty cells (no links) (Corso et al., 2008).

To take into account the effect of the weight of the links on nestedness, from the packed matrix we calculate two weighted adjacency matrices, $P_{ij}^c$ and $P_{ij}^r$ that depict the dependence of column element *j* on row element *i* and vice versa. Thus, the weights of element $M_{ij}$ over row *i* and column *j* are obtained:

$$P_{ij}^c = \frac{M_{ij}}{\sum_{j=1}^{N_r} M_{ij}}, \quad P_{ij}^r = \frac{M_{ij}}{\sum_{i=1}^{N_c} M_{ij}} \qquad (1)$$

where $N_r$ and $N_c$ are the number of rows and columns of the matrix, respectively.

WINE is based on the concept of estimating nestedness through the calculation of a Manhattan distance from each of the matrix cells containing a link to the cell corresponding to the intersection of the row and columns with the lowest marginal totals (number of links). This concept resembles in a way the one used by Corso et al. (2008), although the distances are measured to the opposite corner of the packed matrix. Additionally, in WINE, the Manhattan distance is replaced by a weighted Manhattan distance.

Following Corso et al. (2008), distance properties are introduced in the packed matrix by mapping it into a Cartesian space and rescaling it to the unit square to avoid distortions in asymmetric matrices. Thus, the elements of matrix $(i,j)$ are assigned the positions $x_i$, $y_j$:

$$x_i = \frac{(i-1)}{N_c} + \frac{1}{2N_c}$$
$$y_j = \frac{(j-1)}{N_r} + \frac{1}{2N_r} \qquad (2)$$

With the new positions, we calculate a weighted-interaction distance, which estimates nestedness taking into account the number of events in the links,

$$d_{ij}^w = P_{ij}^r \cdot x_i + P_{ij}^c \cdot y_j \qquad (3)$$

This equation shows the weighted distance of each link and provides a useful parameter for assessing the importance of each link in the network. The greater this distance is, the greater the contribution of this link to nestedness. The weighted-interaction nestedness of the matrix (WIN) is then the mean weighted distance of all its non-zero elements:

$$d^w = \frac{1}{N_l} \sum_{i,j} d_{ij}^w \qquad (4)$$

where $N_l$ is the total number of links. This parameter takes into account both the relative position of links in the matrix and the number of events in each interaction, thus including more precise information than previous nestedness estimators.

The statistical significance of any nestedness index value has to be tested against some null hypothesis. The respective null distributions are obtained from null models that generate expected index values and the associated confidence limits (Ulrich and Gotelli, 2007). In our case, we use a null model that constrains matrix fill to observed values, retains the distribution of number of events in the links but does not constrain marginal totals. By constraining matrix fill and replicating the distribution of number of events we acknowledge that, in a given system, just a fraction of the links are feasible and that the differences in abundance of the interacting elements will condition the distribution of the intensity of the links. However, in terms of marginal totals we follow the equiprobable-equiprobable null model considering that there is no *a priori* reason to assume that certain links are less probable than others (Ulrich and Gotelli, 2007). To generate random matrices, we first calculate the events distribution of links in the original data matrix (Figure 1). Then, we create random matrices with the same size and events distribution of the number of events in the links as in the original data matrix. In these random matrices, we preserve the set of numbers of events of the interactions and the number of interactions to the whole matrix. However, this algorithm does not conserve the number of links for each species. The random matrices are then packed and, subsequently, the weighted-interaction nestedness (WIN) of the random matrices are calculated according to equation 1. We average random nestedness over 100 random matrices to obtain the so-called, $d_{rnd}$. We have chosen 100 random matrices because increasing the number of replicates only delays the calculation of WINE with no significant improvement in accuracy.

We calculate the standardized static variable Z-score, $z = \dfrac{d^w - d_{rnd}}{\sigma}$, to assess the significance of WIN, where $\sigma$ is the standard deviation. This score assesses how different WIN $d^w$ is from the average random WIN $d_{rnd}$. It is important to mention that the original data is just one combination of all possible permutations of registered events. Z values below -1.65 or above 1.65 indicate approximate statistical significance at the 5% error level (one-tailed test).

The WIN of the data matrix can then be normalized by comparing it to the average WIN of equivalent random matrices and to the WIN of the maximal nestedness matrix (Wright and Reeves, 1992) to obtain the weighted-interaction nestedness estimator.

To generate the maximal nestedness matrix, we create a new packed matrix with the same size and number of links. This matrix is constructed so that the links are placed as close as possible to the lowest-rightmost cell in the matrix. Subsequently, the WIN of the maximal nestedness matrix ($d_{max}$) is calculated according to equations 1-4.

Finally, we calculate the weighted-interaction nestedness estimator,

$$\eta_w = \frac{d^w - d_{rnd}}{d_{max} - d_{rnd}} \qquad (5)$$

The value of this estimator approaches zero when the WIN of the original data matrix is close to the average WIN of the equivalent random matrices, and it approaches one as it gets closer to the nestedness of the maximal nestedness matrix. Thus, this estimator evaluates the relative position of the data matrix between the corresponding random matrices and the maximal nestedness matrix. Negatives values for this estimator can be found in some synthetic matrices that have been described as "anti-nestedness" matrices (Table I) (Almeida-Neto et al. 2007). Results in Table I show that synthetic matrices used to evaluate the anti-nestedness models by Almeida-Neto and collaborators have values close to zero or negative values. Negative values indicate that the synthetic matrices are less nested than the corresponding random matrices.

We have tested the new estimator on a range of matrices of various sizes and shapes and observed that results are not affected by these factors. To measure the influence of matrix size and shape, we have calculated the $\eta$ of some datasets and their transposed matrices (Almeida-Neto et al., 2008) and found no significant differences between them. For example, the value of $\eta$ of the transposed of dataset in Figure 1 differs in the third decimal; for (20 x 500)-, (500 x 20)- and (100x100)-random matrices $\eta$ equals to zero; and for the (20 x 500)- and (500 x 20)-checkerboard matrices WINE is -0.06 in both cases. Obviously, as any statistical parameter this estimator is sensitive to small datasets. We have also tested WINE against NTC and NODF with real plant-pollinator matrices. Obviously, as weight distribution is likely to vary from one dataset to other the correlations will depend on the datasets that are used. A set of eight plant-pollinator matrices provided a 0.922 correlation (p<0.001) with NTC and a 0.722 correlation (p<0.023) with NODF.

We have developed a graphical user interface (GUI) running in Matlab, called Weighted-Interaction Nestedness Estimator, WINE (Fig. 2). WINE is a Matlab application developed to perform the calculation of the new weighted-interaction nestedness estimator. This program allows the user to open a data file, select the range of data to be analyzed and calculate the results, obtaining two figures and four indices (see below for definition). The indices shown in the graphical interface are: i) weighted-interaction nestedness (WIN) of the data matrix, ii) average WIN of 100 random matrices of similar characteristics, iii) z-score of WIN with probability $p$ (probability of having the same WIN as a random matrix) and iv) the weighted-interaction nestedness estimator (WINE). The GUI provides a histogram of events distribution of the data matrix and a figure showing a "color plot" of the packed matrix, in such a way that every link is plotted by a color corresponding to the weighted distance. A C++ version of WINE is also available for users. This version is less user-friendly but it can be used in all platforms. It computes the above-mentioned indices but does not currently produce color plots.

We want to emphasize the relevance of the color plot in depicting the relative importance of each interaction, for instance, in the identification of idiosyncratic species or in the evaluation of extinction sequences in species distribution in fragmented habitats. In Figure 3 we depict two color plots of weighted distances showing the difference between the packed original dataset (left) and a packed random dataset created with the same events distribution (right). The dataset corresponds to an interaction matrix. Dark red bands correspond to links of high relative importance in an extinction sequence when dealing with species distribution in fragmented habitats.

In conclusion, this new estimator has all the advantages of the nestedness index proposed by Corso et al. (2008), overcoming the limitations of some previous indices: it is independent from matrix size and fill (if it is not almost empty). In addition to taking into account the relative position of the links in the matrix, it accounts for the number of events of links, i.e. weighted interactions, through $d^w$, thus including more information than previous estimators. This estimator overcomes some of the artifacts derived of using different sampling intensities because weights are probabilities. Thus, different sampling intensities will change absolute values of registered events, $M_{ij}$, but weights will be invariant as long as a representative range of sampling intensities is used. This is true as long as higher sampling intensity does not discover new links and lower sampling intensity does not eliminate any existing link or distort frequencies due to lack of power.

We believe that this new estimator can be of interest to the scientific community, and especially to those involved in the study of biotic interactions in ecology. Barrat et al. (2004) showed that heterogeneity in the intensity of interactions may be very important in the understanding and full description of social and biological networks. To comply with this need, our new weighted-interaction nestedness estimator $\eta$ not only contains the standard topological information shown in the presence-absence matrix, but also adds information on the intensity of interactions, which is computed to estimate the nestedness of the matrix.

Acknowledgements

We are grateful to A. Traveset and J.M. Olesen for testing the program, providing matrix data and asking questions that helped improve WINE software. We thank three anonymous reviewers for their helpful suggestions for improving the manuscript.We also thank Lori J. De Hond for her linguistic assistance. This research was partially funded by project EXTREM (CGL2006-09431) of the Spanish Ministry of Education and Culture.

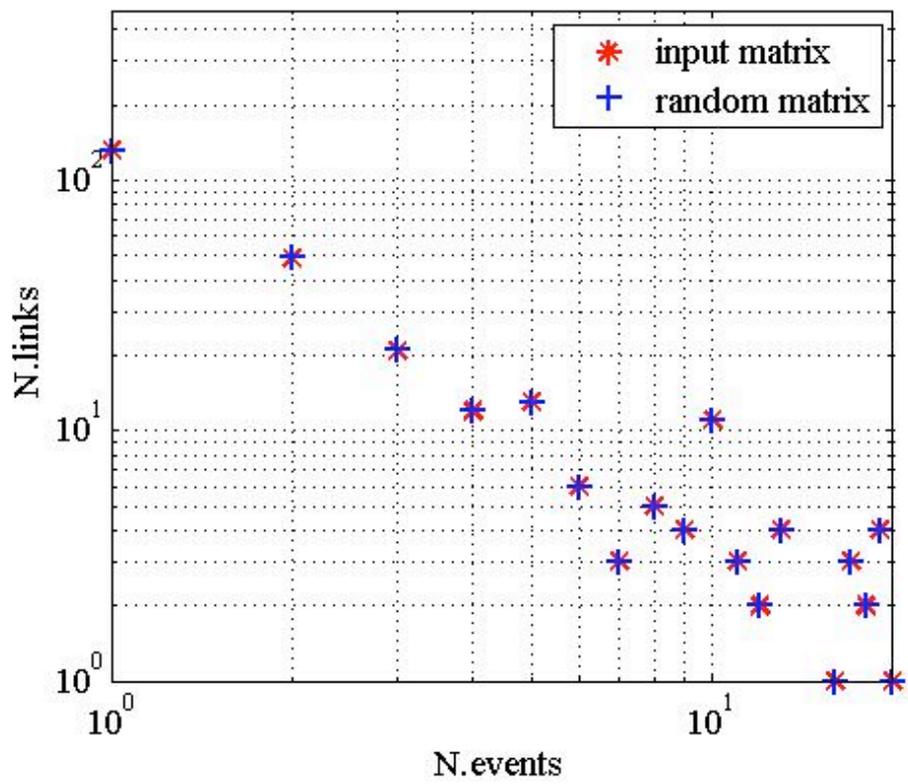

Figure 1. Frequency distribution of number of events in the input matrix (circles) and in simulated random matrices (plus signs).

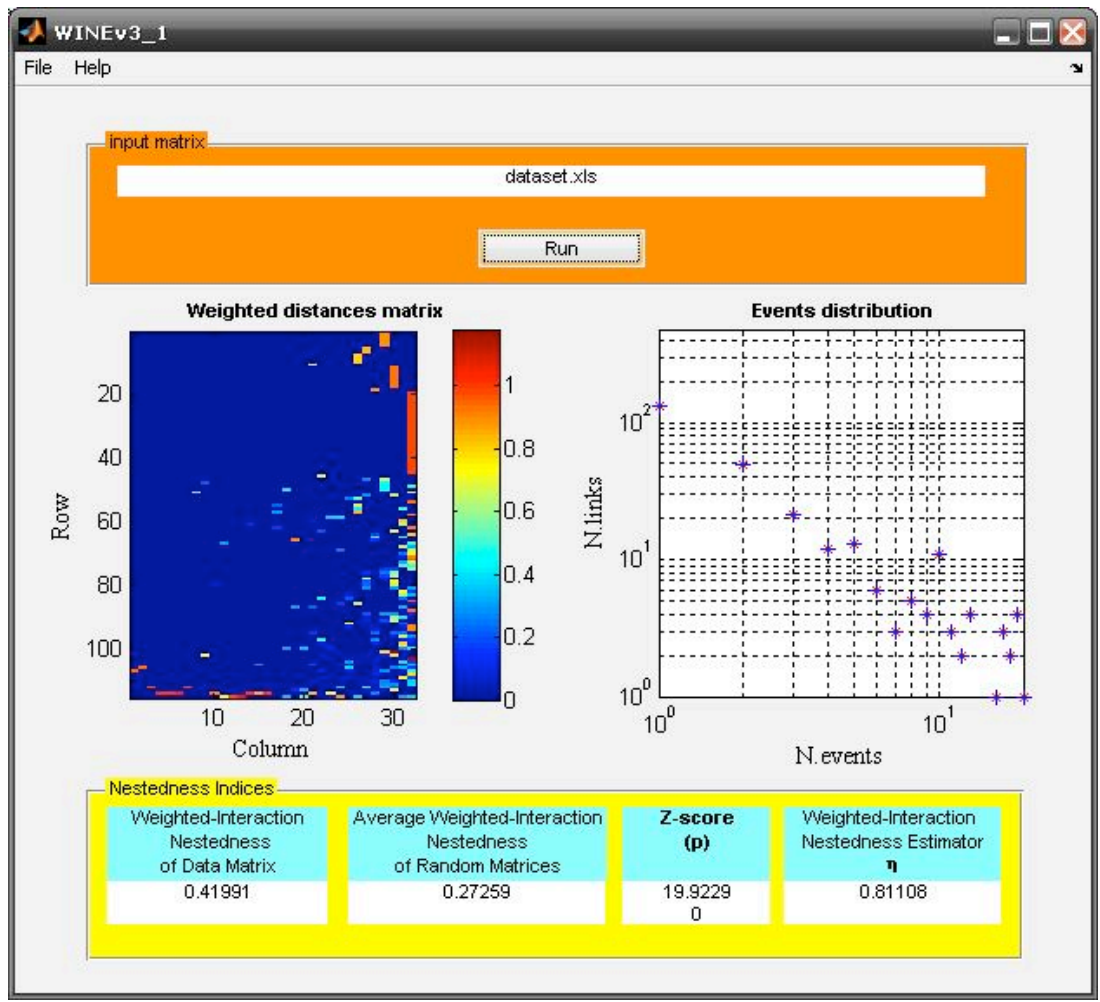

Figure 2. Output of WINE after running a dataset file. The figure shows the graphical interface with two plots: the weighted distance of each link on the left and the histogram of events distribution on the right. WIN values of the data matrix, average WIN of the random matrices, z-score and WINE are provided in the lower part.

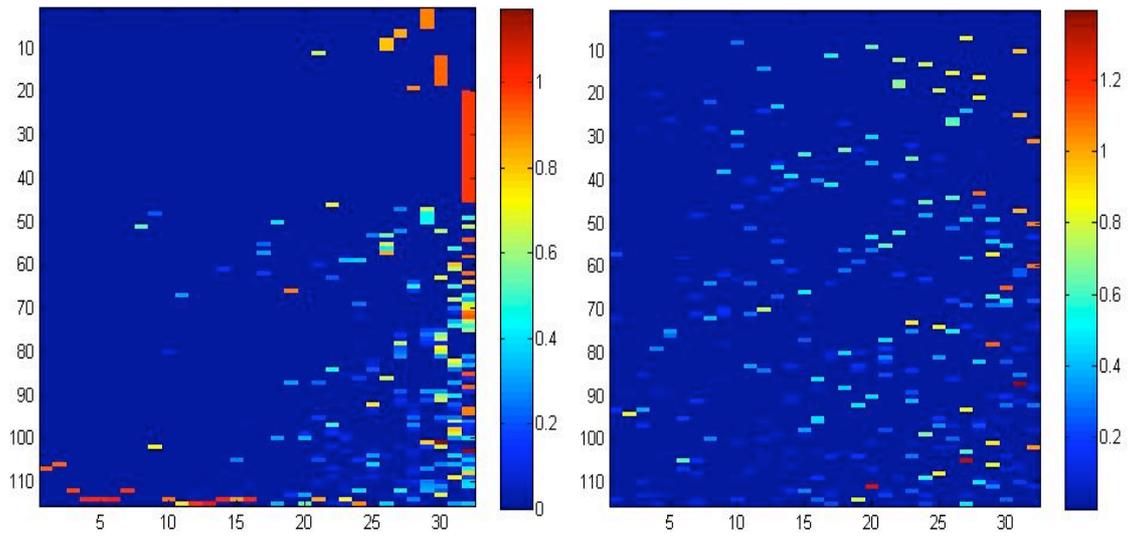

Figure 3. Weighted distance plots of an interaction dataset (left) and its packed random matrix, with the same events distribution (right).

| Matrix | WINE (η) | z | P |
|---|---|---|---|
| Checkerboard (15 x 6) (fill=50%) | -0.4 | -2.4 | 0.01 |
| Compartmented (15 x 6) (fill=33%) | -0.3 | -2.1 | 0.02 |
| High_turnover (15 x 5) (fill=73%) | -0.25 | -1,4 | 0.08 |
| Random (15 x 6) (fill=50%) | 0.08 | 0.5 | 0.33 |

Table I. Nestedness analysis of different synthetic matrices used by Almeida-Neto et al. (2007, 2008) to discuss the nestedness and "anti-nestedness" concepts (checkerboard, compartmented, high-turnover and random). $\eta$: WINE estimator, $z$: z-score of synthetic matrices, $p$: probability of the matrix having a WIN-value, $d^w$, less than expected by chance, $d_{rnd}$, according to the null model.